# Channel Capacity Analysis of MIMO System in Correlated Nakagami-m Fading Environment

Samarendra Nath Sur[#1], Dr. Rabindranath Bera[#2] and Dr. Bansibadan Maji[*3]

[#]*Department of Electronics and Communication Engineering,*
*Sikkim Manipal Institute of Technology, Sikkim, India*
[*]*Department of Electronics and Communication Engineering,*
*National Institute of Technology, Durgapur, West Bengal, India*

*Abstract—* **We consider Vertical Bell Laboratories Layered Space-Time (V-BLAST) systems in correlated multiple-input multiple-output (MIMO) Nakagami-m fading channels with equal power allocated to each transmit antenna and also we consider that the channel state information (CSI) is available only at the receiver. Now for practical application, study of the VBLAST MIMO system in correlated environment is necessary. In this paper, we present a detailed study of the channel capacity in correlated and uncorrelated channel condition and also validated the result with appropriate mathematical relation.**

## I. INTRODUCTION

Multiple Input Multiple Output (MIMO) systems have established the potential for increased capacity by exploiting the spatial properties of the multi-path, without requiring additional bandwidth [1]. Lots of research has been carried out towards the enhancement of spectral efficiency and the solution that came out as the practical model to obtain large spectral efficiency leading to the Bell Laboratories Layered space–time (BLAST) architecture [2].

It was shown in [1] that MIMO channels, whereby multiple antennas are employed at both the transmitter and receiver, offer large gains in capacity over single-input–single-output (SISO) channels [3] and also one of the important techniques used to exploit the spatial diversity in a rich scattering environment in order to improve the spectral efficiency [4].The capacity increases linearly with the number of antennas, provided that the channel gain coefficients are independent identically distributed complex Gaussian variables [5]. Because of this highly required potential, from the perspective of the today's wireless communication scenario, MIMO technology has been introduced in 802.16, 3GP project for mobile communication. However, in a real world scenario, spatial correlation occurs due to poor scattering or limited angular spread and insufficient antenna spacing. Therefore, sub channels between the pairs of transmit-receive antennas are not independent; rather those sub channels are spatially correlated. This can cause the channel capacity of MIMO systems reduce significantly. In case of a fully correlated MIMO system, it virtually provides one sub channel, which means no advantage is provided by a MIMO system [6].

In wireless communication scenario, due to the presence of the physical object in the signal propagation path, such as trees, buildings, the signals get polluted by some physical phenomenon such as absorption, reflection, refraction, diffraction, and scattering etc. [7]. Channel fading effect severely degrades the communication system performance. More distinctively removal of fast fading effect is more decisive for the system performance and can be solved by exploiting the diversity technique. Lots of research work has been carried out to study the behavior of wireless communication systems over different fading models specially Rayleigh and Rician fading. However, in this paper, we have studied the Nakagami distribution as Nakagami-m fading channel represents various fading condition in wireless channel [7] [8]. This wide range of distribution can be realized by varying fading index m. It becomes the Rayleigh distribution when m = 1 and for m > 1 the Rician fading can be closely approximated by Nakagami-m fading model. Also, it becomes the one-sided Gaussian distribution (m → 0.5), and Nakagami-m distribution covers no fading channel as m goes to infinity [9] [10]. So analysis of Nakagami channel is very important.

In the case of outdoor rich scattering environment, due to the presence of scatters there will be correlated fading at the both ends. Based on this, we consider the MIMO single user case with correlated receive and correlated transmit antennas. And also we have analyzed the MIMO system performance based on the channel capacity with the linear receiver such as ZF, MMSE at the receiver side.

## II. MATHEMATICAL MODEL

We consider a MIMO antenna system with $N_t$ transmitting antenna and $N_r$ receiving antennas. The received signal y can be represented by

$$Y = Hx + n \qquad \text{----- (1)}$$

Where the transmit symbols vector x satisfies, and $E\{xx^\tau\} = Q$ and n is the $N_r \times 1$ additive white Gaussian noise vector. The vector H represents the slowly varying flat fading channels for the wireless transmission. The channel is assumed to be independent and identically distributed (i.i.d) and which follows a Nakagami –m fading probability distribution function (pdf). Let γ represent the instantaneous SNR and for Nakagami fading channel it can be defined as which can be defined [11] by

$$P_\gamma(\gamma) = \frac{2}{\Gamma(m)} \left(\frac{m}{\bar{\gamma}}\right)^m \gamma^{m-1} \exp\left(\frac{-m\gamma}{\bar{\gamma}}\right), \quad \gamma \geq 0 \quad \text{-----(2)}$$

As in [12] the sub-channel capacity of a MIMO system with a ZF linear receiver can be defined as

$$C_{ZF} = \log_2\left[1 + \frac{SNR}{(H^H H)^{-1}}\right] \quad \text{-----(3)}$$

Now from [13], we have

$$H^H H = (R_{RX}^{1/2} G R_{TX}^{H/2})^H \cdot (R_{RX}^{1/2} G R_{TX}^{H/2})$$
$$= R_{RX}^{H/2} G^H R_{TX}^{1/2} R_{RX}^{1/2} G R_{TX}^{H/2}$$
$$= R_{RX} G G^H R_{TX} \quad \text{-----(4)}$$

Where $R_{TX}$ and $R_{RX}$ represents the correlation matrices for the transmitter and the receiver side. The spatial correlation between the transmit antennas ($R_{TX}$) is assumed to be independent from the correlation between the receive antennas ($R_{RX}$). Here G is independent and identically distributed complex Gaussian zero mean unit variance elements. Therefore $C_{ZF}$ can be written as

$$C_{ZF} = \log_2\left[1 + \frac{SNR}{(G^H G R_{RX} R_{TX})^{-1}}\right]$$
$$= \log_2\left[1 + SNR(G^H G R_{RX} R_{TX})\right] \quad \text{-----(5)}$$

The sub-channel capacity of a MIMO system with MMSE linear receiver [12] can be defined as

$$C_{MMSE} = \log_2\left[\frac{1}{(I_{N_r} + SNR \cdot H^H H)^{-1}}\right]$$

For high SNR

$$(I_{N_r} + SNR \cdot H^H H)^{-1}$$
$$= (SNR \cdot H^H H)^{-1} - I_{N_r}(SNR \cdot H^H H)^{-2} + O(SNR^{-3})$$
$$= (SNR \cdot H^H H)^{-1}\left[1 - I_{N_r}(SNR \cdot H^H H)^{-1} + O(SNR^{-2})\right]$$

Therefore,

$$C_{MMSE} = \log_2\left[\frac{SNR \cdot H^H H}{\left[1 - I_{N_r}(SNR \cdot H^H H)^{-1} + O(SNR^{-2})\right]}\right]$$

$$C_{MMSE} = \log_2\left[\frac{SNR \cdot R_{RX} G G^H R_{TX}}{\left[1 - I_{N_r}(SNR \cdot R_{RX} G G^H R_{TX})^{-1} + O(SNR^{-2})\right]}\right] \quad \text{-----(6)}$$

The capacity difference of a MIMO system with MMSE and ZF receiver can be written as

$$C_{MMSE} - C_{ZF}$$
$$= \log_2(SNR) - \log_2\left[1 - I_{N_r}(SNR \cdot R_{RX} G G^H R_{TX})^{-1} + O(SNR^{-2})\right]$$
$$- \log_2\left[SNR + (R_{RX} G G^H R_{TX})^{-1}\right] \quad \text{-----(7)}$$

Now, analyze the effect of the channel correlation on the MIMO system with the MMSE and ZF receiver.

The Channel capacity of a MIMO system with ZF receiver in uncorrelated and correlated channel condition is given below.

$$C_{ZF_{UNCor}} = \log_2\left[1 + \frac{SNR}{(G^H G)^{-1}}\right] = \log_2\left[1 + SNR \cdot (G^H G)\right] \quad \text{----(8)}$$

$$C_{ZF_{Cor}} = \log_2\left[1 + SNR(G^H G R_{RX} R_{TX})\right] \quad \text{----(9)}$$

Now $\log(1+x) = x - \frac{x^2}{2} + \frac{x^3}{3} - \ldots\ldots\infty$

Using the above relation equations (8) and (9) can be expressed as given below

$$C_{ZF_{UNCor}} = \left[SNR(G^H G)\right]\left[1 - \frac{SNR(G^H G)}{2} + O(SNR^2)\ldots\infty\right] \quad \text{----(10)}$$

$$C_{ZF_{Cor}} = \left[SNR(G^H G R_{RX} R_{TX})\right] \times$$
$$\left[1 - \frac{SNR(G^H G R_{RX} R_{TX})}{2} + O(SNR^2)\ldots\ldots\infty\right] \quad \text{----(11)}$$

Therefore,

$$\frac{C_{ZF_{Cor}}}{C_{ZF_{UNCor}}} = \frac{R_{RX} R_{TX}\left[1 - \frac{SNR(G^H G R_{RX} R_{TX})}{2} + O(SNR^2)\ldots\infty\right]}{\left[1 - \frac{SNR(G^H G)}{2} + O(SNR^2)\ldots\ldots\infty\right]} \quad \text{----(12)}$$

= 1 for very low correlated channel

< 1 for fully correlated channel.

The Channel capacity of a MIMO system with MMSE receiver in uncorrelated and correlated channel condition is given below.

For large SNR

$$C_{MMSE_{UNCor}} = \log_2 \left[ SNR \cdot (G^H G) \right]$$
$$- \log_2 \left[ 1 - I_N (SNR \cdot (G^H G))^{-1} + O(SNR^2) \right] \quad \text{---- (13)}$$

$$C_{MMSE_{Cor}} = \log_2 \left[ SNR (G^H G R_{RX} R_{TX}) \right]$$
$$- \log_2 \left[ 1 - I_N (SNR (G^H G R_{RX} R_{TX}))^{-1} + O(SNR^2) \right] \quad \text{---- (14)}$$

$$C_{MMSE_{UNCor}} - C_{MMSE_{Cor}} = \log_2 \left[ \frac{SNR(G^H G)}{SNR(G^H G R_{RX} R_{TX})} \right]$$
$$+ \log_2 \left[ \frac{1 - I_N (SNR(G^H G R_{RX} R_{TX}))^{-1} + O(SNR^2)}{1 - I_N (SNR \cdot (G^H G))^{-1} + O(SNR^2)} \right]$$

Now, neglecting the higher order SNR terms, we have,

$$C_{MMSE_{UNCor}} - C_{MMSE_{Cor}} = \log_2 \left[ \frac{1}{R_{RX} R_{TX}} \right]$$
$$+ \log_2 \left[ \frac{(SNR(G^H G R_{RX} R_{TX})) - I_N}{R_{RX} R_{TX} \{SNR \cdot (G^H G) - I_N\}} \right]$$
$$= \log_2 \left[ \frac{1}{(R_{RX} R_{TX})^2} \cdot \frac{(SNR(G^H G R_{RX} R_{TX})) - I_N}{\{SNR \cdot (G^H G) - I_N\}} \right] \quad \text{-----(15)}$$

### III. SIMULATION RESULTS

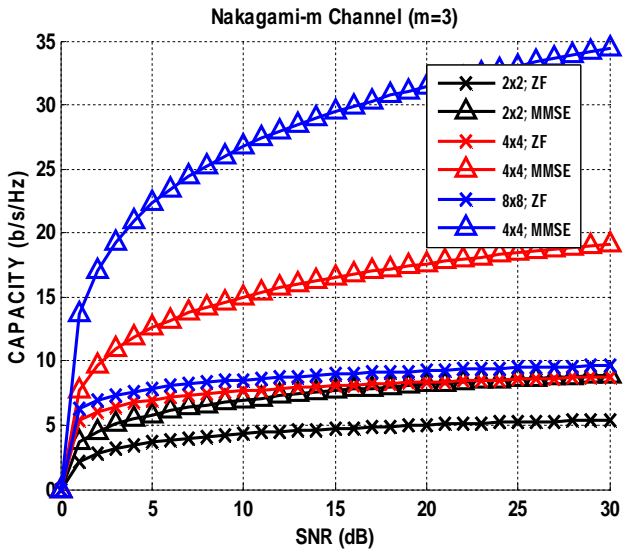

Fig. 1. Channel capacity vs. SNR curves in the Nakagami channel (m=3) with MMSE and ZF receiver.

Above figure 1 depicts the channel capacity variation with SNR for un-correlated Nakagani-m channel (m=3) with MMSE and ZF receiver. The figure shows the improvements in the channel capacity with the increase in the number of antennas in the transmitter and receiver side and also the capacity difference for MMSE receiver and ZF receiver increases with the increase in the diversity order.

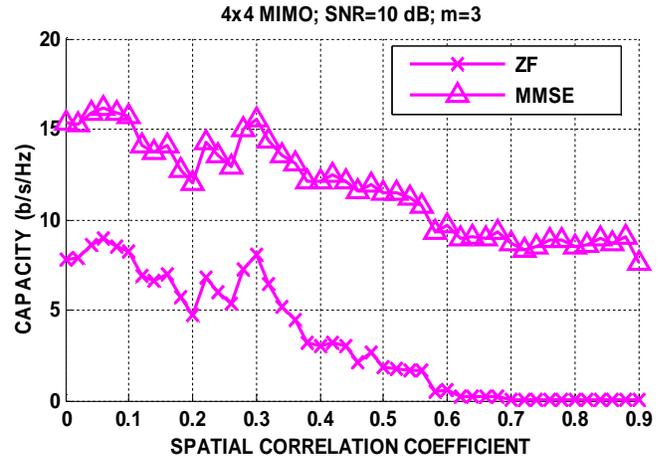

Fig. 2. Channel capacity vs. Spatial correlation coefficient curves in the Nakagami channel (m=1) with MMSE and ZF receiver.

Figure 2 shows the variation in channel capacity with the spatial correlation coefficient for a 4 X 4 MIMO system with MMSE and ZF receiver. Here we consider BPSK digital modulation technique with the spatial correlation on the both transmitter and receiver side. As in figure we observed that with the increase in the spatial correlation coefficient from 0 (un-correlated) to 1 (fully correlated), the channel capacity decreases significantly. And we also observed a significant degradation in performance of the MIMO system with ZF receiver for higher correlated channel in comparison to that of an MMSE receiver.

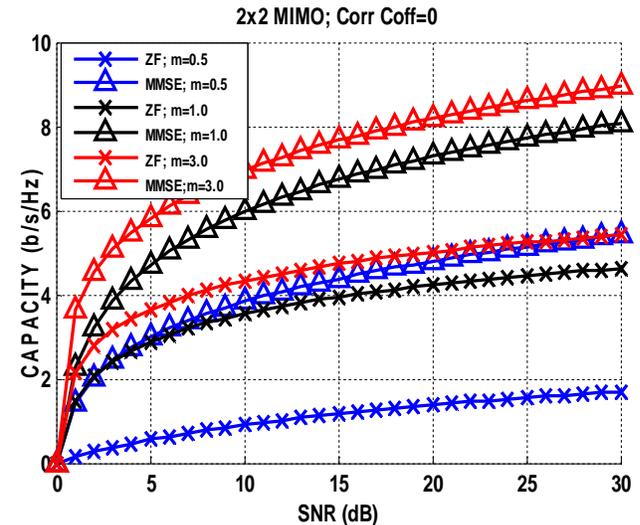

Fig. 3. Channel capacity vs. SNR curves in the Nakagami channel with MMSE and ZF receiver.

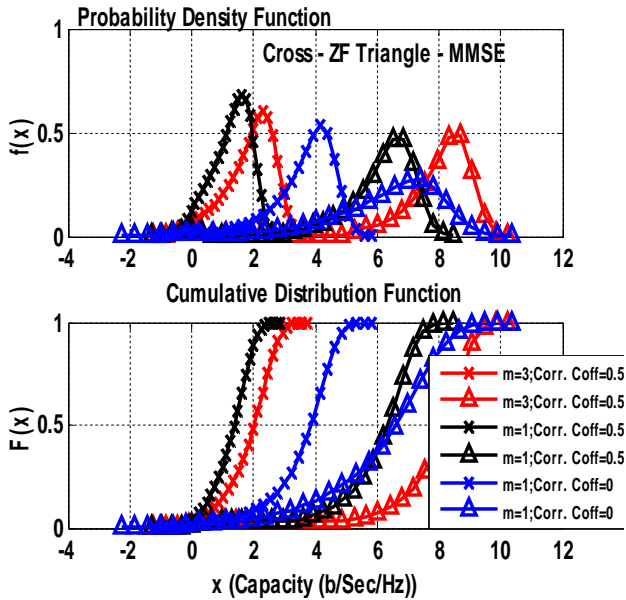

Fig. 4. PDF and CDF of Channel capacity of 2x2 MIMO system in correlated Nakagami channel with MMSE and ZF receiver.

Figure 3 and 4 depict the performance comparison of a MIMO system in correlated Nakagami channel with MMSE and ZF receiver. The influence of Nakagami fading parameter, m and correlated channel is represented in the above said figures.

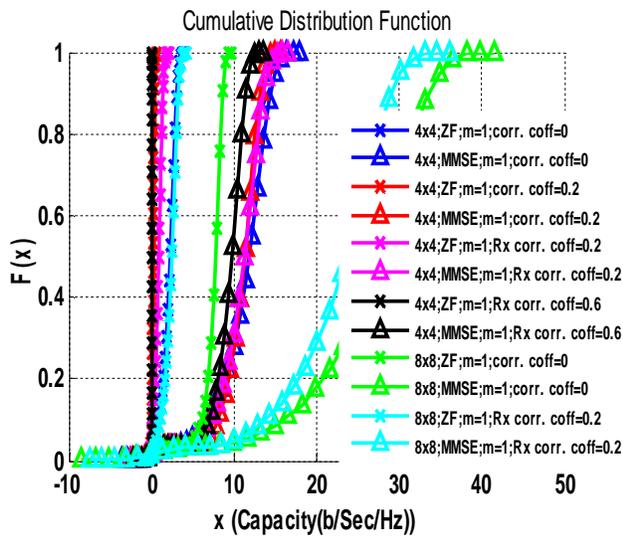

Fig. 5a. CDF of Channel capacity of MIMO system for correlated Nakagami channel with MMSE and ZF receiver.

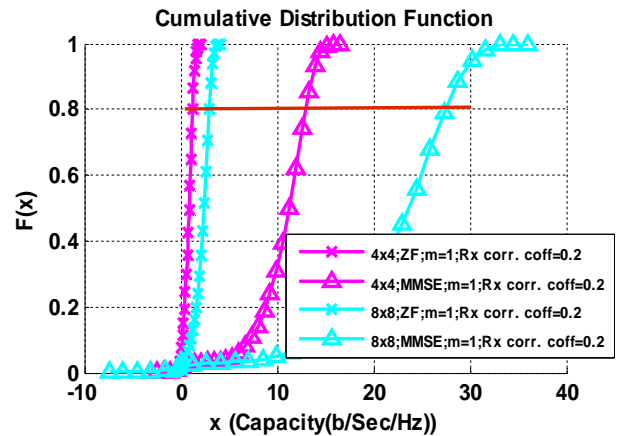

Fig. 5b. Comparison between 4x4 and 8x8 MIMO in correlated Nakagami channel with MMSE and ZF receiver.

The statistical analysis of the channel capacity is represented in figures 4 and 5. From figure 5b, at $F(x) = 0.8$, the observed channel capacity values are represented in the table given below.

TABLE I
CAPACITY COMPARISON

| Observations | System Specification | Capacity (b/Sec/Hz) |
|---|---|---|
| 1 | 4x4 MIMO; ZF receiver; m=1; Rx Corr Coff=0.2 | 1.26 |
| 2 | 8x8 MIMO; ZF receiver; m=1; Rx Corr Coff=0.2 | 2.975 |
| 3 | 4x4 MIMO; MMSE receiver; m=1; Rx Corr Coff=0.2 | 12.92 |
| 4 | 8x8 MIMO; MMSE receiver; m=1; Rx Corr Coff=0.2 | 27.32 |

From the table and the represented curves it is clear that in correlated channel, MMSE is the best linear receiver in comparison to its counterpart ZF receiver.

## IV. CONCLUSIONS

Analysis of the correlated MIMO channel is indeed an important issue from the real environmental application perspective. This paper deals with the detail analysis of the spatial correlation effect on the MIMO system with MMSE and ZF receiver. Here we find that the spatial correlation at the transmitter and receiver side puts limitation over the system spectral efficiency. From the above presented mathematical calculation and the simulation result, we find that in the correlated channel MMSE receiver provide significantly better performance in comparison to that of ZF receiver.

REFERENCES

[1] G. Foschini, "On limits of wireless communication in fading environment when using multiple antennas," Wireless Personal Communication, no.6, pp.311–335, Mar.1998.
[2] G. J. Foschini, "Layered space-time architecture for wireless communication in a fading environment when using multiple antennas," Bell Labs Tech.J., vol.1, no.2, pp.41–59, Autumn 1996.
[3] Geoffrey J.Byers, and Fambirai Takawira, "Spatially and Temporally Correlated MIMO Channels: Modeling and Capacity Analysis", IEEE Transactions On Vehicular Technology, Vol.53, No.3, May 2004, Pp-634-643.
[4] Jraifi Abdelouahed1 and El Hassan Saidi, "Optimization of MIMO Systems in a Correlated Channel", IJCSNS, VOL.8 No.2, February 2008, pp-277-282.
[5] Telatar I., "Capacity of Multiantenna Gaussian Channels," Trans-European Telecommunication Networks, vol. 10, no. 6, pp. 585-595, 1999.
[6] Jun Wang, Quan Zhou, Jinfang Dou, Lede Qiu, "On the Channel Capacity of MIMO Systems under Correlated Rayleigh Fading" WiCom, 2007, pp 134-136.
[7] Li Tang and Zhu Hongbo, "Analysis and Simulation of Nakagami Fading Channel with MATLAB", CEEM-2003, China, Nov. 4-7,2003. pp. 490-494.
[8] M. Nakagami, "The m-distribution, a general formula of intensity distribution of rapid fading," in Statistical Methods in Radio Wave Propagation, W. G. Hoffman, Ed, Oxford, England: Pergamum, 1960.
[9] George K.Karagiannidis, Dimitris A.Zogas, and Stavros A. Kotsopoulos "On the Multivariate Nakagami-m Distribution With Exponential Correlation", IEEE Transactions On Communications, Vol.51, No.8, August2003.
[10] A. Annamalai, and C. Tellambura, "Error Rates for Nakagami-m Fading Multi-channel Reception of Binary and M-ary Signals", IEEE Transactions On Communications, Vol. 49, No.1, January 2001.
[11] Hyundong Shin, and JaeHong Lee, "On the Error Probability of Binary and M-ary Signals in Nakagami-m Fading Channels" IEEE Transactions On Communications, Vol.52, No.4, April2004.
[12] Onggosanusi, Eko N.; Dabak, Anand G.; Schmidl, Timothy; Muharemovic, Tarik; "Capacity analysis of frequency-selective MIMO channels with sub-optimal detectors", ICASSP, 2002.
[13] Duong, T.Q.; Hoang Trang; Hong, E.K.; Lee, S.Y., "Performance evaluation of the V-BLAST system under correlated fading channels", AICT, 2005.